\documentclass[twocolumn]{aastex631}
\usepackage{float}
\usepackage{graphicx}
\usepackage[T1]{fontenc}
\usepackage{aecompl}
\usepackage[]{amsmath}
\usepackage{color}
\usepackage{xspace}
\usepackage{soul}
\usepackage{booktabs} 
\usepackage{array}     
\usepackage{multirow} 

\shorttitle{$n-M_h$ relations}
\shortauthors{Liu et al.}
\graphicspath{{./}{figures/}}

\begin{document}

\title{The Dependence of Dark Matter Halo Properties on the Morphology of Their Central Galaxies from Weak Lensing}

\correspondingauthor{Zhenjie Liu, Kun Xu}
\email{liuzhj26@sjtu.edu.cn, kunxu@sas.upenn.edu}

\author[0009-0000-3649-5365]{Zhenjie Liu}
\affiliation{State Key Laboratory of Dark Matter Physics, School of Physics and Astronomy, Shanghai Jiao Tong University, Shanghai 200240, China}
\affiliation{Division of Physics and Astrophysical Science, Graduate School of Science, Nagoya University, Nagoya 464-8602, Japan}

\author[0000-0002-7697-3306]{Kun Xu}
\affiliation{Center for Particle Cosmology, Department of Physics and Astronomy,
University of Pennsylvania, Philadelphia, PA 19104, USA}
\affiliation{State Key Laboratory of Dark Matter Physics, School of Physics and Astronomy, Shanghai Jiao Tong University, Shanghai 200240, China}
\affiliation{Institute for Computational Cosmology, Department of Physics, Durham University, South Road, Durham DH1 3LE, UK}

\author{Jun Zhang}
\affiliation{State Key Laboratory of Dark Matter Physics, School of Physics and Astronomy, Shanghai Jiao Tong University, Shanghai 200240, China}

\author{Wenting Wang}
\affiliation{State Key Laboratory of Dark Matter Physics, School of Physics and Astronomy, Shanghai Jiao Tong University, Shanghai 200240, China}

\author{Cong Liu}
\affiliation{State Key Laboratory of Dark Matter Physics, School of Physics and Astronomy, Shanghai Jiao Tong University, Shanghai 200240, China}

\begin{abstract}
Xu \& Jing reported a monotonic relationship between host halo mass $M_h$ and the morphology of massive central galaxies, characterized by the S\'ersic index $n$, at fixed stellar mass, suggesting that morphology could serve as a good secondary proxy for halo mass. Since their results were derived using the indirect abundance matching method, we further investigate the connection between halo properties and central galaxy morphology using weak gravitational lensing. We apply galaxy-galaxy lensing to measure the excess surface density around CMASS central galaxies with stellar masses in the range of $11.3 < \log M_*/{\rm M_\odot} < 11.7$, using the HSC shear catalog processed through the Fourier\_Quad pipeline. By dividing the sample based on $n$, we confirm a positive correlation between $n$ and $M_h$, and observe a possible evidence of the positive correlation of $n$ and halo concentration. After accounting for color, we find that neither color nor morphology alone can determine halo mass, suggesting that a combination of both may serve as a better secondary proxy. In comparison to hydrodynamic simulations, we find that TNG300 produce much weaker correlations between $M_h$ and $n$. Furthermore, using SIMBA simulations with different feedback mode, we find jet-mode active galactic nuclei feedback might be related to the relationship of S\'ersic index and halo mass.
\end{abstract}


\section{Introduction}

In the $\Lambda$CDM cosmological model, the large-scale structure of the universe is shaped by the gravitational influence of dark matter, such as halos, filaments and voids. These dark matter halos subsequently act as ``seeds'' for galaxy formation. Dark matter, a substance that has not yet been directly observed, constitute the major mass of the universe, whereas stars are the luminous objects we can observe. The Stellar-to-Halo Mass Relation (SHMR) reveals the intimate connection between the stellar mass of galaxies and the mass of dark matter halos, indicating the link between galaxy formation and the evolution of dark matter halos. It is generally believed that massive dark matter halos host more massive galaxies. Therefore, by utilizing SHMR, we can assign galaxy masses to halos in N-body simulations, or infer halo mass from stellar mass. This allows for a deeper understanding of the evolutionary history of galaxies, and their coevolution with dark matter halos.

Although SHMR provides critical insights into the connection between galaxies and halos, it also exhibits significant scatter \citep{2010MNRAS.402.1942C,2023ApJ...944..200X,2014MNRAS.443.3044Z,2021MNRAS.505.5117Z,2022MNRAS.511.1789Z}, potentially correlated with the influence of other galaxy properties such as size, color, and morphology. For instance, some studies find that halos of blue (star-forming) galaxies are less massive than those of red (quenched/passive) galaxies with the same stellar mass \citep{2015ApJ...799..130R,2016MNRAS.457.3200M,taylor2020gama+,wang2021stellar,xu2022photometric}. Conversely, other studies \citep{tinker2013evolution,Moster2018,guo2019evolution} report the opposite conclusion. 

The morphology of galaxies is also found to correlate with halo mass. Galaxy morphology is typically described by the S\'{e}rsic profile \citep{sersic1963influence}, which characterizes the brightness, size, and the morphology of a galaxy. The S\'{e}rsic profile is modeled as:
\begin{equation}\label{eq:sersic}
 I(r) = I_e {\rm exp}\bigg \{ -b_n \bigg[ \bigg(\frac{R}{R_e}\bigg)^{\frac{1}{n}} - 1 \bigg] \bigg \} ,
\end{equation}
where $b_n$ is defined through $\gamma(2n;b_n)=\Gamma(2n)$, $\Gamma$ and $\gamma$ are the Gamma function and the lower incomplete Gamma function, respectively. $I_e$ is the surface brightness at the half-light radius $R_e$. The S\'{e}rsic index $n$ quantifies the curvature of the galaxy's luminosity profile, with higher values indicating more concentrated core of light distributions. In this work, we focus on the relation between halo masses and the morphology of galaxies, characterized by the S\'ersic index (\(n\)). Studies from \cite{sonnenfeld2019hyper} utilize lensing data from the Hyper Suprime-Cam Survey (HSC) to study the relationship between halo mass and stellar mass, galaxy size, and S\'{e}rsic index in the CMASS sample. They find no significant correlation between halo mass and size or S\'{e}rsic index at a fixed stellar mass. However, \cite{taylor2020gama+} find that for galaxies with a stellar mass of around $10^{10.5}{\rm M_\odot}$, the S\'{e}rsic index and effective radius are the best indicators of host halo mass, while showing weaker correlations with star formation rate or color. Applying the Photometric Objects Around Cosmic Webs \citep[PAC;][]{2022ApJ...925...31X} method, \cite{xu2022photometric} find that more compact, red, and larger galaxies at fixed stellar mass tend to be located in more massive halos. But their method of estimating halo mass relies on abundance matching, which assumes a one-to-one correspondence between stellar mass and the satellite distribution of the host halo. Therefore, the relationship between galaxy morphology and halo mass remains unclear, and the physical origin influencing the SHMR also remains an open question.

Weak gravitational lensing can serve as a crucial probe to the total mass of dark matter halos and the matter distribution, which distorts background galaxy images around massive objects, known as shear. Galaxy-galaxy lensing, in particular, can directly measure the average density profile of halos around galaxies or galaxy clusters, thus reflecting the halo mass. In this work, we utilize the lensing data from the third public data release of HSC to further analyze and discuss the dependence of the host halo mass on the morphology of massive central galaxies with stellar masses in the range of $11.3 <  {\rm log}(M_*/{\rm M_\odot}) < 11.7$. We introduce our galaxy sample and shear catalog in Section \ref{sec:data}, and describe our lensing measurement methods in Section \ref{sec:measurement}. In Section \ref{sec:results}, we analyze the measured lensing signals and the corresponding dependence on central galaxy morphologies and the properties of the galaxy clusters\footnote{We made our results public on https://github.com/load666/arXiv-2411-07525/}. Subsequently, in Section \ref{sec:siml}, we explore the underlying physical origins using the SIMBA and TNG simulations. Finally, we summarize our findings in Section \ref{sec:conclusion}.

\section{Data}\label{sec:data}

\subsection{CMASS}
We use the galaxy morphology catalog measured by \cite{xu2022photometric}, with the galaxy sample selected from Baryon Oscillation Spectroscopic Survey (BOSS, \cite{ahn2012ninth,bolton2012spectral}) constant-mass (CMASS), with an area of approximately 500 deg$^2$ overlapping with HSC. They identify galaxies that do not have more massive neighbors within a projected distance of 1 Mpc/\(h\), considering these to be central galaxies. However, they did not account for fiber collisions or the stellar mass completeness of CMASS \citep{2016MNRAS.457.4021L,2018ApJ...858...30G,2023ApJ...944..200X}, both of which can bias the selection of the most massive galaxies and lead to off-centering effects. In our lensing model, we incorporate these factors, along with the possibility that central galaxies may not be the most massive and that there may be offsets between the centers of halos and their central galaxies \citep{Hikage2013, Wang2014, Lange2017}. In their work, stellar mass and rest-frame colors are measured by fitting the spectral energy distribution. The S\'{e}rsic indices are obtained by fitting the S\'{e}rsic profile to the corresponding HSC z-band images, as detailed in \cite{xu2022photometric}. In this work, we use galaxies with stellar masses in the narrow range of \(11.3 < \log (M_*/{\rm M_\odot}) < 11.7\), a redshift range of 0.45-0.7, and S\'{e}rsic indices ranging from 0.8 to 8.

\subsection{HSC}\label{sec:HSC}

The shear catalog we use is processed through the Fourier\_Quad \citep[FQ;][]{2008MNRAS.383..113Z} pipeline, based on the third public data release of HSC, which covers an area of about 1400 deg$^2$ and includes about 100 million galaxies \citep{Liu2024zlh}. The FQ method employs multipole moments of galaxy images in Fourier space for shear measurements, including five shear estimators: \(G_1\), \(G_2\), \(N\), \(U\), and \(V\). \(G_i\) is based on the galaxy quadrupole moments in Fourier space. $N$ is a normalization factor used to estimate and correct the impact of noise and the Point Spread Function (PSF). $U$ and $V$ are the additional terms taking care of the parity properties of the shear estimators in the PDF-SYM method, which is a novel way of estimating shear statistics. Their specific definitions can be found in \cite{Zhang2017}. 

The shear catalog records basic information about galaxies such as position, signal-to-noise ratio ($\nu_F$, \cite{Li2021}), magnitude, and shear estimators. Photometric redshift (photo-$z$) information is sourced from the DEmP method in \cite{nishizawa2020photometric}, which demonstrates very low bias and scatter for photo-z less than 1.5. Therefore, we limit the maximum photo-$z$ of background galaxies to 1.5. Additionally, since the scatter photo-$z$ for our sample is estimated to be about 0.02-0.04 (Table 2 in \cite{nishizawa2020photometric}), to reduce the intrusion of foreground galaxies due to photo-$z$ errors,  we use the background galaxies whose photo-$z$ are larger than the lens redshift plus 0.1, i.e., \(z_s > z_l + 0.1\).  To enhance the accuracy of shear measurements, we select background galaxies with signal-to-noise ratio larger than 4 and absolute field distortion $|g_f|$ smaller than 0.02 (see Appendix \ref{sec:FD} for more details on $g_f$). HSC observes in five bands: $g$, $r$, $i$, $z$, and $y$, with the $i$ band offering the best imaging quality. In the FQ method, shear estimators are measured for each galaxy image without stacking images from different bands. Following \cite{Liu2024zlh}, we choose data from the $r$, $i$, and $z$ bands as the shear sample, as their multiplicative shear bias in the field distortion tests is at about 1-2 percent level at most, and their galaxy-galaxy lensing measurements show consistency with the results from the HSC year-one official shear catalog. 

More recently, it is found in \cite{shen2024} that additional shear multiplicative bias can be induced by the photo-z cut of the background sample, which is a selection effect unknown before. We therefore find it necessary to calibrate for the multiplicative bias that may arise due to the selection of a subset of the background sample in the measurement of, e.g., galaxy-galaxy lensing. Fortunately, the FQ shear catalog keeps the field distortion information for each galaxy image/shear estimator, it therefore provides a convenient on-site calibration approach. Its details relevant to this work is provided in Appendix \ref{sec:FD}.   
 
 \subsection{Simulations}
In our work, we study two hydrodynamic simulations, TNG and SIMBA, to compare our observational results and explore the origins of various relationships. TNG is a series of large cosmological gravity-magnetohydrodynamics simulations that incorporate comprehensive models of galaxy formation physics \citep{PillepichSimulating2017}. We utilize the simulation with a box length of $L_{\rm box} = 205  {\rm Mpc}/h$, denoted as TNG300. The SIMBA simulation provides detailed modeling of black hole growth and feedback, successfully reproducing many observational results \citep{dave2019simba, thomas2019black, cui2021origin}. Specifically, AGN feedback in SIMBA operates in three modes based on the accretion rate: the radiative mode at high Eddington ratios $f_{\rm Edd}$, and the jet and X-ray modes at low $f_{\rm Edd}$. For our primary analysis, we use the full physics simulation with $L_{\rm box} = 100 \, {\rm Mpc}/h$, denoted as S100. According to \cite{WeinbergerSupermassive2018, dave2019simba, Montero2019Mergers}, kinetic feedback modes are primarily responsible for quenching massive central galaxies in both SIMBA and TNG, while major galaxy mergers are not the main drivers of most quenching events. Thus, we also utilize various AGN feedback simulations from SIMBA with $L_{\rm box} = 50 \, {\rm Mpc}/h$ (S50) to study their effects.
 
We first select central galaxies with stellar masses in the range of  $11.3 < {\rm log}M_*/{\rm M_\odot} < 11.7$ from TNG300 and S100 to match the observation, and then choose the range
$11 < {\rm log}M_*/{\rm M_\odot} < 11.5$ for S50 series simulations due to their limited volume. The stellar mass is defined as the sum of the masses of stars within twice the stellar half-mass radius. Assuming a constant mass-to-light ratio, we create galaxy images for each galaxy's stellar mass distribution, extending to $4 \times 4$ times the half-stellar mass radius. Then we convolve a small gaussian PSF to reduce noises. We fit these images with a 2D ${\rm S \Acute{e}rsic}$ profile to obtain their ${\rm S \Acute{e}rsic}$ index $n$. Additionally, we fit the surface density of the host halos with an NFW profile to obtain their halo mass $M_h$ and concentration $c$. The definitions in NFW model are aligned with our lensing measurements. We use the position with the minimum gravitational potential energy as the real center of halos, hence the off-centering effect can be ignored in this part.

\section{Measurement}\label{sec:measurement}
\subsection{PDF-SYM for ESD}\label{sec:PDF}
Galaxy-galaxy lensing measures the correlation between the position of an object and the surrounding shear, which can reflect the excess surface density (ESD) around the object. For a spherically symmetric lensing potential, the relationship between shear and ESD is given by
\begin{equation}\label{eq:ESD}
\Delta \Sigma (R)=\overline{\Sigma} (<R)-\Sigma(R)=\Sigma_c \gamma_t(R),
\end{equation}
where $\overline{\Sigma}(<r)$ refers to the average surface density within a radius $r$. $\Sigma_c$ is the comoving critical surface density, defined as 
\begin{equation}\label{sigmac}
    \Sigma_c=\frac{c^2}{4 \pi G } \frac{D_{\rm s}}{D_{\rm l} D_{\rm l s}(1+z_{\rm l})^2}
\end{equation} 
Here, $c$ is the speed of light, $G$ is the gravitational constant, and $D_{\rm s}, D_{\rm l}$, and $D_{\rm ls}$ are the angular diameter distances for the lens, source, and lens-source systems, respectively.

In our work, we employ the PDF-symmetrization (PDF-SYM, \cite{Zhang2017}) method to measure the ESD of galaxies. The essence of this method lies in constructing the Probability Density Function (PDF) of the shear estimators, and adjusting the assumed underlying shear value to achieve the most symmetric state of the PDF, thereby obtaining the optimal shear estimate. For the ESD signal at a given radius $r$, we can obtain the PDF of the shear estimators $P(G_t)$ for the background galaxies. 
By hypothesizing an ESD signal $\widehat{\Delta \Sigma}$, we adjust $P(G_t)$ to $P(\hat{G}_t)$ by modifying each value of $G_t$ via:
\begin{equation}\label{Ghat}
\hat{G}_t=G_t-\frac{\widehat{\Delta \Sigma}}{\Sigma_c}\left(N \pm U_t\right),
\end{equation}
where the $N$ and $U_t$ is the normalization factor and rotated correction factor introduced in Sec.\ref{sec:HSC}. When \( \widehat{\Delta \Sigma} \) equals the true ESD signal, the new distribution \( P(\hat{G}_t) \) will achieve the most symmetric state.

In this study, we uniformly divide the radius range from 0.05 to 50 Mpc/$h$ into 9 bins in logarithmic space for measurement and estimate the covariance matrix by dividing the sky into 86 sub-regions using Jackknife method. Additionally, to remove the effects of boundary from the observed region and some systematic bias, we subtract  the signal contribution from random points, the number of which is 10 times the number of lens galaxies. Additionally, we correct the multiplicative biases introduced in shear measurements and details are shown in Appendix \ref{sec:FD}.

\subsection{Modeling and Fitting}\label{sec:model}
For the ESD model, we take into account the effects of stellar mass of galaxies, individual halos, off-centering effects, and the 2-halo term. 

In this work, we assume that the galaxy luminosity distribution follows a S\'{e}rsic profile. However, relative to halos, the mass of galaxies is concentrated within a very small radius, allowing it to be treated as a point source for its ESD and the ESD of galaxy is
\begin{equation}\label{eq:ESD_Ms}
\Delta  \Sigma_*(R)=\frac{M_*}{\pi R^2}
\end{equation}
where $M_*$ is the stellar mass of central galaxy, as shown in Table \ref{tab:baseres} for all subsamples. We briefly verify the point-source model in Appendix \ref{app:sersic}, showing that its ESD is consistent with that of the S\'{e}rsic profile within our radius range.

We adopt NFW model \citep{navarro1997universal} to describe an individual halo profile, whose density is expressed as
\begin{equation}\label{eq:rhor}
\rho (r)=\frac{\rho_0}{(r/r_s)(1+r/r_s)^2}
\end{equation}
with $\rho_0=\rho_m \Delta_{\rm vir}/(3I)$, where $I=[\ln (1+c)-c/(1+c)]/c^3$. $r_{\rm vir}$ is the virial radius, $r_s$ is the scale radius and the concentration $c$ of halos is defined by $c= r_{\rm vir}/r_s$. In this paper, we define a halo as a sphere with an overdensity of \( \Delta_{\rm vir} = 180 \) times the mean matter density \( \rho_m \), with a halo mass $M_h=180 (4 \pi /3) \rho_m r^3_{\rm vir}$. The unit of $M_h$ in this paper is ${\rm M_\odot}/h$. We employ the analytical expression for the surface density \( \Delta \Sigma_{\rm NFW} \) as presented in \cite{yang2006}.

The off-centering effect is a significant consideration within the virial radius of halos, as it can lead to an underestimation of the ESD signal at smaller radii. Ignoring this effect might result in an underestimation of the halo's mass and concentration. Thus, the 1-halo term ESD model typically divides the halo profile into well-centered and off-centered components,
\begin{equation}\label{sigma1h}
\Delta \Sigma_{\rm 1h} (R)=f_{\rm c}\Delta \Sigma_{\rm NFW}(R) +(1-f_{\rm c})\Delta \Sigma_{\rm off}\left(R \right) 
\end{equation}
where $f_{\rm c}$ is the proportion of well-centered halos. $\Sigma_{\rm off}$ refers to mean surface density of mis-centered halos, which can be derived from
\begin{equation}\label{eq:sigmaoff}
\begin{aligned}
\Sigma_{\text {off }}\left(R \right)&=\frac{1}{2 \pi} \int_0^{\infty} {\rm d} R_{\rm off} P\left(R_{\rm off} \right)   \int_0^{2 \pi} {\rm d} \theta\\
 &\times \Sigma_{\rm NFW} \left(\sqrt{R^2+R_{\rm off}^2+2 R R_{\rm off} \cos \theta}\right)  .
\end{aligned}
\end{equation}
where $P\left(R_{\rm off} \right)$ is the distribution of offset centers on radius and it is generally assumed to be Rayleigh distribution \citep{offcen2007}
\begin{equation}\label{Prs}
P\left(R_{\rm off} \right)=\frac{R_{\rm off}}{\sigma_s^2} \exp \left(-\frac{R_{\rm off}^2}{2 \sigma_s^2}\right)
\end{equation}
$\sigma_s$ represents the scatter of off-centered positions. In this work, we set it as a free parameter and use the halo radius $r_{\rm vir}$ as the unit.

Given that our measurement radii significantly exceed the common virial radii of halos, the 2-halo term exerts a substantial influence on the ESD. Following \cite{van2013} and \cite{Wang2022}, the surface density of the 2-halo term can be expressed as,
\begin{equation}\label{sigma2h}
 \Sigma_{\rm 2h}(R)=2 \rho_m \int_R^\infty [1+ \xi_{\rm gm}^{\rm 2h}(r)]\frac{r {\rm d}r}{\sqrt{r^2- R^2}}
\end{equation}
where $\xi_{\rm gm}^{\rm 2h}(r)= b_h \zeta (r,z) f_{\rm exc}(r) \xi_{\rm mm}(r)$. $b_h$ is the halo bias, which is a free parameter in our model. $\zeta (r,z)$ is radial bias function obtained in \cite{van2013},
\begin{equation}
\zeta(r, z) = 
\begin{cases} 
\zeta_0(r, z) & \text{if } r \geq r_{\psi} \\ 
\zeta_0(r_{\psi}, z) & \text{if } r < r_{\psi}
\end{cases}
\end{equation}
where
\begin{equation}
\zeta_0(r, z) = \frac{[1 + 1.17\xi_{\text{mm}}(r, z)]^{1.49}}{[1 + 0.69\xi_{\text{mm}}(r, z)]^{2.09}}
\end{equation}
and
\begin{equation}
    \log \left[ \zeta_0 (r_\psi, z) \, \xi_{\text{mm}} (r_\psi, z) \right] = 0.9.
\end{equation}
$\xi_{\rm mm}(r)$ is calculated by COLOSSUS \citep{diemer2018colossus} using a semi-analytical power spectrum from \cite{Eisenstein1998}. $f_{\rm exc}(r)$ describes the halo exclusion effect, where it equals 0 when $r<r_{\rm vir}$ and equals 1 otherwise.

Combining Eqs.\ref{eq:ESD_Ms}-\ref{sigma2h} and Eq.\ref{eq:ESD}
, our ESD model can eventually be summarized as
\begin{equation}\label{sigmatot}
\Delta \Sigma (R)=\Delta  \Sigma_*(R)+\Delta  \Sigma_{\rm 1h}(R| M_h, c,\sigma_s, f_c)+\Delta  \Sigma_{\rm 2h}(R | b_h)
\end{equation}
To enhance the accuracy of our model, we divide the redshift range of 0.45 to 0.7 into five bins and perform an integration based on the distribution of lens galaxies. Although more sophisticated modeling approaches can be adopted to model the scatter of halo mass or others \citep[e.g.][]{Leauthaud2012ApJ, Han2015MNRAS}, we leave more sophisticated modeling to future studies. Also, we test our model in simulations shown later.

In total, our model comprises five free parameters, halo mass $M_h$, concentration $c$, halo bias $b_h$, the scatter of off-centering $\sigma_s$ and the proportion of well-centered halos $f_c$, which we fit using Markov Chain Monte Carlo techniques \citep{Christensen_2001}. The flat priors of the parameters in the fitting process are listed in Table \ref{tab:prior}.

\begin{table} 
\centering  
\caption{The prior of all fitting parameters in MCMC approach.} \label{tab:prior}
\begin{tabular}{l|ccccc}
\hline  
 Parameter & log$M_h$& $c$& $\sigma_s$ & $f_c$ & $b_h$\\
[2.4pt]
\hline
Prior & [12.5,14.5] &[0,20] & [0.01,1.2] & [0,1] &[0,10] \\
\hline
\end{tabular}  
\end{table} 

\section{Results and discussion}\label{sec:results}

\subsection{The relationship between halo properties and S\'{e}rsic index}
\begin{figure*}[ht!]
\centering
\includegraphics[width=0.7\textwidth]{ 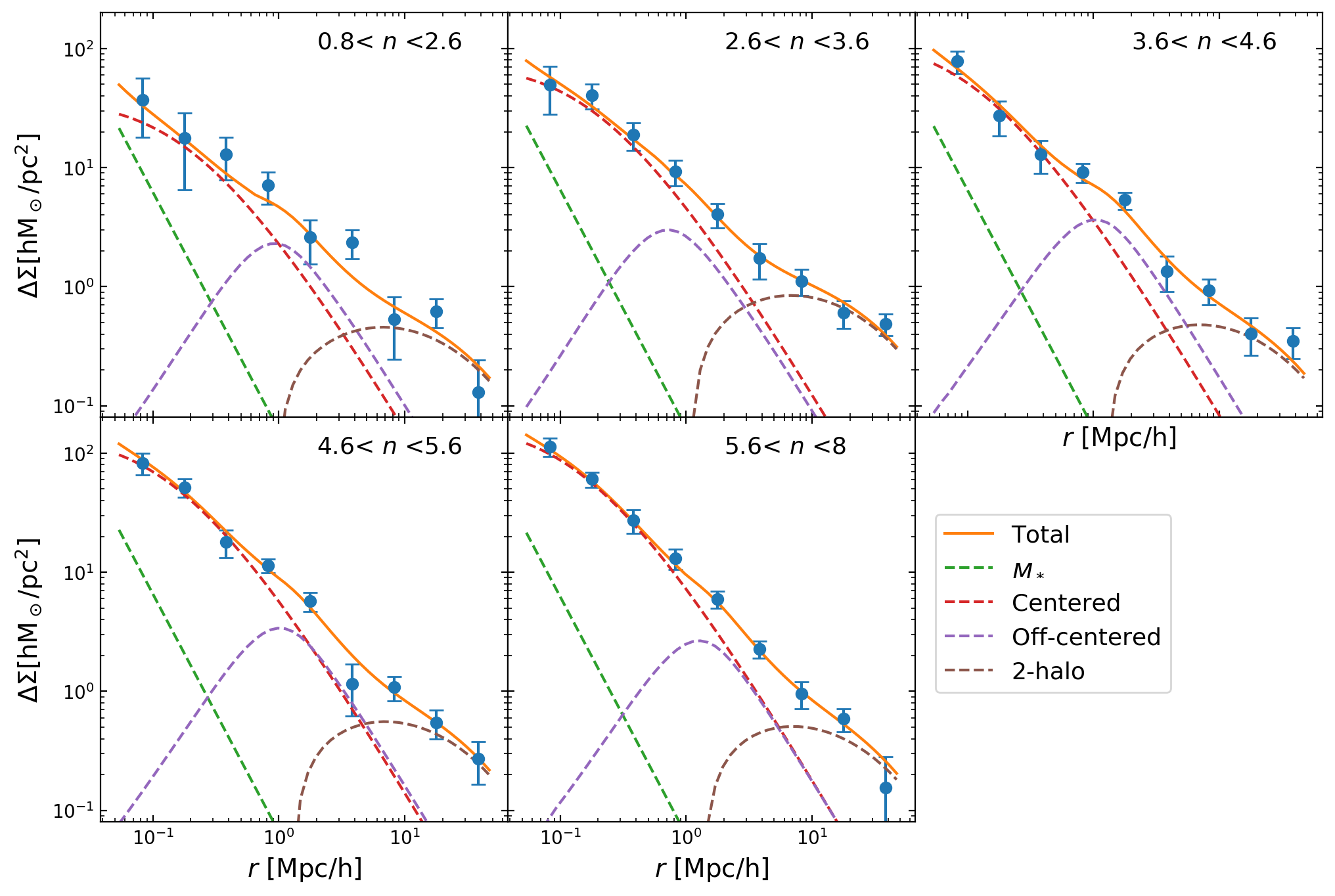}
\caption{Measurement of the lensing ESD around central galaxies with different ${\rm S \Acute{e}rsic}$ indices $n$. Solid lines represent the best-fit curves, and dashed lines represent ESD of different components, including stellar mass, well-centered portion, off-centered portion and 2-halo term.\label{fig:esd}}
\end{figure*}

\begin{figure*}[ht!]
\centering
\includegraphics[width=1\textwidth]{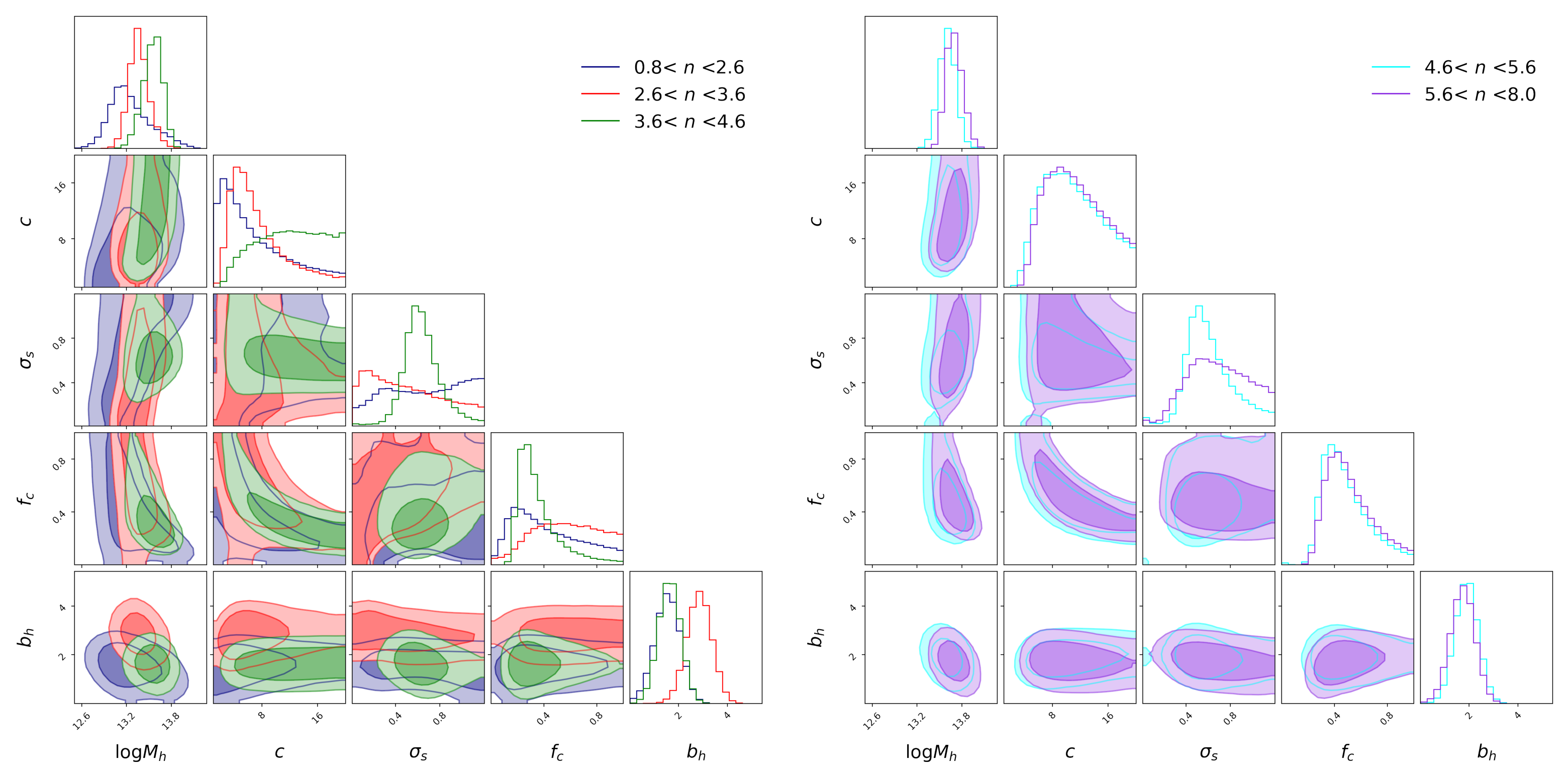}
\caption{The 68\% and 95\% confidence level contour plots of 5 free parameters for ``All'' samples with different S\'ersic indices.} \label{fig:contour}
\end{figure*}

\begin{figure*}[ht!]
\centering
\includegraphics[width=1.2\columnwidth]{ 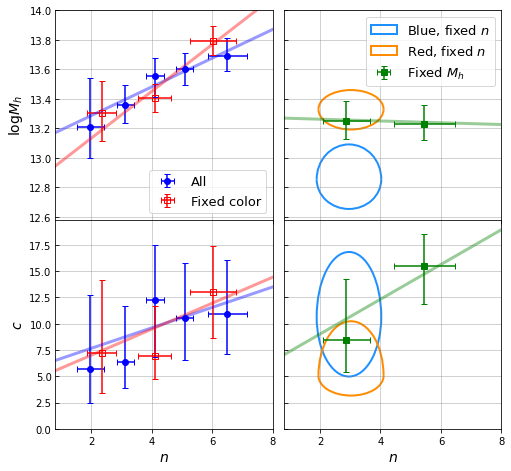}
\caption{Best-fit halo mass ${\rm log}M_h$ and concentration $c$ under different conditions. Blue dots represent all samples with different ${\rm S \Acute{e}rsic}$ indices $n$, and red dots represent results for different $n$ with fixed sample color. Light blue and orange rings represent results for blue and red galaxies, respectively, while fixed $n$, with ring size representing the 1$\sigma$ range. \label{fig:nmc}}
\end{figure*}

\begin{table*} 
\centering  
\caption{Fitting results for halo mass ${\rm log}M_h$, concentration $c$, scatter in off-centering $\sigma_s$, the proportion of well-centered galaxies $f_c$ and halo bias $b_h$ in different subsamples. The results in the parentheses are the ratios of the measured bias $b_h$ to the theoretical value $b_h(M_h,z)$ modeled by \cite{tinker2010large}, with the measured halo mass $M_h$ and mean redshift of samples. ``All'' refers to samples controlled only by constant stellar mass. ``Fixed color'' refers to samples where galaxy color is also controlled, and ``Fixed $n$'' refers to samples controlled by ${\rm S \Acute{e}rsic}$ index. ``Fixed $M_h$'' is to control the the halo mass of subsamples by applying the selection on $M_G$, where $M_G$ is the halo mass assigned by abundance matching in group catalog \cite{yang2021extended}.
Middle panel shows conditions of subsamples and the information, including the number of galaxies $N_{\rm lens}$ and their average stellar mass log$M_*/{\rm M_\odot}$.} 
\label{tab:baseres}
\begin{tabular}{c|ccc|cccccc}
\hline
 Samples & $n$ / color & $N_{\rm lens}$& log$M_*$& ${\rm log}M_h$ & $c$ &$\sigma_s$ & $f_c$ & $b_h$ & $\chi^2/\nu$ \\
[2.4pt]
\hline
 &$0.8<n<2.6$ &1453&11.46 &
 $13.21^{+0.33}_{-0.21}$ & 
$5.67^{+7.02}_{-3.24}$ & 
$0.68^{+0.37}_{-0.41}$ & 
$0.37^{+0.36}_{-0.21}$ & 
$1.53^{+0.51}_{-0.53} (0.8^{+0.3}_{-0.3})$ &  2.41
\\
 [2.4pt]
&$2.6<n<3.6$ &1744&11.48& 
$13.36^{+0.14}_{-0.12}$ & 
$6.34^{+5.34}_{-2.50}$ & 
$0.46^{+0.44}_{-0.31}$ & 
$0.57^{+0.28}_{-0.26}$ & 
$2.84^{+0.50}_{-0.52} (1.4^{+0.3}_{-0.3})$ &
0.94
 \\
 [2.4pt]
All  &$3.6<n<4.6$& 1985&11.48 &  
$13.56^{+0.12}_{-0.14}$ & 
$12.26^{+5.20}_{-5.44}$ & 
$0.62^{+0.17}_{-0.13}$ & 
$0.31^{+0.19}_{-0.09}$ & 
$1.62^{+0.45}_{-0.45} (0.7^{+0.2}_{-0.2})$ & 
2.17
  \\
 [2.4pt]
& $4.6<n<5.6$& 1883&11.48&  
$13.60^{+0.11}_{-0.11}$ & 
$10.49^{+5.25}_{-3.96}$ & 
$0.58^{+0.26}_{-0.15}$ & 
$0.45^{+0.21}_{-0.13}$ & 
$1.89^{+0.47}_{-0.47} (0.8^{+0.2}_{-0.2})$ &  
1.72
  \\
 [2.4pt]
& $5.6<n<8$&1783&11.46&  
$13.69^{+0.12}_{-0.10}$ & 
$10.96^{+5.06}_{-3.87}$ & 
$0.68^{+0.31}_{-0.27}$ & 
$0.48^{+0.23}_{-0.14}$ & 
$1.72^{+0.49}_{-0.50} (0.7^{+0.2}_{-0.2})$ &  
0.40
  \\
 [2.4pt]
\hline
 &$0.8<n<3$ &826&11.50 &
 $13.31^{+0.21}_{-0.19}$ & 
$7.19^{+6.94}_{-3.78}$ & 
$0.51^{+0.48}_{-0.30}$ & 
$0.48^{+0.32}_{-0.25}$ & 
$1.10^{+0.59}_{-0.59} (0.6^{+0.3}_{-0.3})$ & 
1.37
  \\
 [2.4pt]
 Fixed color & $3<n<5$&2560 &11.49& 
 $13.40^{+0.10}_{-0.09}$ & 
$6.92^{+4.78}_{-2.16}$ & 
$0.49^{+0.36}_{-0.27}$ & 
$0.63^{+0.24}_{-0.22}$ & 
$1.99^{+0.36}_{-0.36} (1.0^{+0.2}_{-0.2})$ & 
2.35
  \\
 [2.4pt]
 ($2.3<u-r<2.7$)&$5<n<8$ &2102&11.47 & 
 $13.79^{+0.10}_{-0.10}$ & 
$13.02^{+4.38}_{-4.40}$ & 
$0.66^{+0.25}_{-0.16}$ & 
$0.35^{+0.15}_{-0.08}$ & 
$1.26^{+0.41}_{-0.42} (0.5^{+0.1}_{-0.2})$ &  
1.29
 \\
 [2.4pt]
\hline
 Fixed $n$ & $2<u-r<2.25$&1195&11.42&  
 $12.85^{+0.24}_{-0.20}$ & 
$10.63^{+6.18}_{-5.66}$ & 
$0.64^{+0.38}_{-0.39}$ & 
$0.51^{+0.30}_{-0.27}$ & 
$2.47^{+0.61}_{-0.62} (1.7^{+0.4}_{-0.4})$ & 
2.73
 \\
 [2.4pt]
 ($1<n<6$) & $2.25<u-r<2.6$&1202&11.43&  
 $13.33^{+0.13}_{-0.14}$ & 
$5.03^{+5.20}_{-1.87}$ & 
$0.26^{+0.54}_{-0.15}$ & 
$0.53^{+0.32}_{-0.32}$ & 
$2.75^{+0.56}_{-0.57} (1.4^{+0.3}_{-0.3})$ & 
1.53
 \\
 [2.4pt]
\hline
 Fixed $M_h$ & $0.8<n<4$&1361&11.49&  
 $13.25^{+0.14}_{-0.12}$ & 
$8.47^{+5.77}_{-3.10}$ & 
$0.36^{+0.59}_{-0.25}$ & 
$0.58^{+0.28}_{-0.30}$ & 
$2.93^{+0.55}_{-0.55} (1.6^{+0.3}_{-0.3})$ & 
1.74
 \\
 [2.4pt]
 & $4<n<8$&1923&11.49&  
$13.23^{+0.13}_{-0.11}$ & 
$15.45^{+3.06}_{-3.62}$ & 
$0.62^{+0.39}_{-0.44}$ & 
$0.77^{+0.16}_{-0.19}$ & 
$1.64^{+0.52}_{-0.52} (0.9^{+0.3}_{-0.3})$ &  0.77
 \\
 [2.4pt]
\hline
\end{tabular}  
\end{table*}

We divide the galaxy samples with stellar masses in the range of $11.3 < {\rm log}M_* /{\rm M_\odot} < 11.7$ into five subsamples. The range of their S\'{e}rsic indices $n$, the number of galaxies, and the average stellar mass are presented in Table \ref{tab:baseres}. Figure \ref{fig:esd} shows the measurements of ESD, with their corresponding best-fit curves and the different components of the model, and Figure \ref{fig:contour} shows their 68\% and 95 \% confidence level contour plots. It is evident that the ESD increases with higher $n$. The ``All'' section in Table \ref{tab:baseres} presents the fitting results for all parameters. To clearly show the trend of the parameters, we depict the relationship between halo mass $M_h$ and halo concentration $c$ as blue dots in Figure \ref{fig:nmc}, with the horizontal axis representing the average $n$ for each sample. We find that halo mass increases significantly with the increase of the ${\rm S \Acute{e}rsic}$ index, which is consistent with the conclusions from \cite{xu2022photometric}. 

Additionally, we observe a possible evidence that more concentrated galaxies (high $n$) may reside in more concentrated halos (high $c$), although the uncertainties are relatively large with current data. We fit the blue dots in the figure using linear relationships
, yielding:
\begin{equation}\label{eq:kxm_all}
{\rm log}M_h=0.10_{-0.04}^{+0.04} n + 13.09_{-0.21}^{+0.22}\quad[{\rm All}],
\end{equation}
\begin{equation}\label{eq:kxc_all}
c=1.0_{-1.4}^{+1.4} n + 5.7_{-5.8}^{+6.6}\quad[{\rm All}].
\end{equation}
Nevertheless, due to the large uncertainties and the strong degeneracy between the concentration parameter $c$ and the off-centering fraction $f_c$, as shown in Figure \ref{fig:contour}, higher-quality data are required to validate our findings. Our best-fit results suggest that only about 50\% of galaxies in our sample reside at the true centers of their halos, consistent with previous weak lensing studies using analyses \citep[e.g.][]{Luo2018ApJ, Wang2022}. In contrast, some theoretical studies based on semi-analytical and hydrodynamical models report lower off-centering fractions \citep[e.g.][]{Planck2013, Wang2019}. Further investigation is necessary to reconcile these differences and confirm the observational results.

Assembly bias refers to the phenomenon where the clustering of dark matter halos is influenced not only by their mass but also by their formation history and surrounding environment \citep{Wechsler2006ApJ, Dalal2008ApJ}. This means that halos of the same mass, but with different formation processes or environments, could exhibit different halo bias. We also set the halo bias $b_h$ as a free parameter in the hope of detecting assembly bias, and show the ratios of the measured bias $b_h$ to the theoretical value $b_h(M_h,z)$ modeled by \cite{tinker2010large} in Table \ref{tab:baseres}. But for subsamples of ``All'', we do not observe a clear trend of $b_h$ or assembly bias varying with $n$.

\begin{figure*}[ht!]
\centering
\includegraphics[width=0.88\textwidth]{ 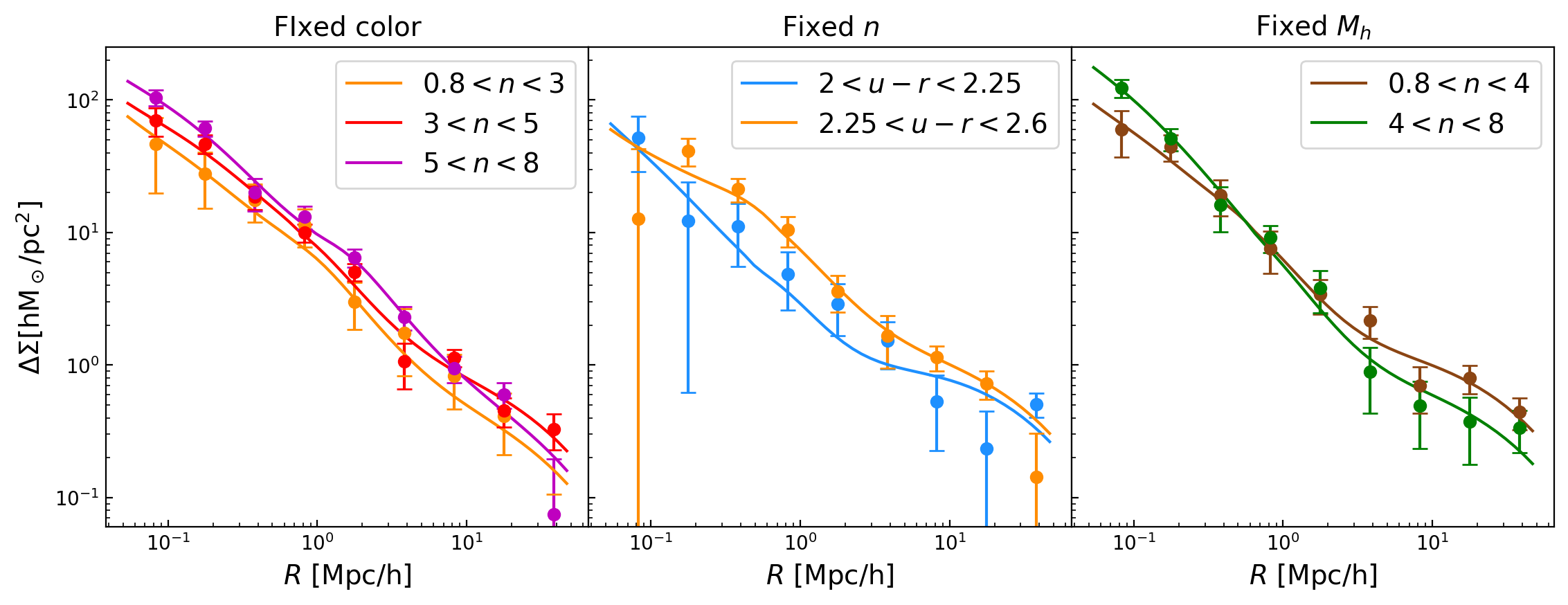}
\caption{Dots with errorbars are measured lensing ESD signals. Solid lines are the best fits. In the left, middle and right panels, galaxies are grouped according to their S\'{e}rsic indices at fixed color, by color at fixed S\'{e}rsic index and by S\'{e}rsic indices at fixed host halo mass obtained through abundance matching. These are indicated by different colors (see the legend). }
\label{fig:fixed}
\end{figure*}

\subsection{The effect of galaxy color and morphology}
In terms of galaxy properties, higher S\'{e}rsic indices are typically associated with early type galaxies with red colors, while lower S\'{e}rsic indices refer to late type galaxies with blue colors \citep[see, e.g., Figure~1 of][]{xu2022photometric}. It has long been recognized that red and blue galaxies exhibit distinct SHMR \citep[e.g.][]{2011MNRAS.410..210M,2012ApJ...757....4P,2012MNRAS.424.2574W,2015ApJ...799..130R,2016MNRAS.457.3200M,2021ApJ...919...25W,2021A&A...649A.119P,xu2022photometric,2023ApJ...947...19A}, with red galaxies hosted by more massive dark matter halos than blue ones at fixed stellar mass. Given the strong correlation between morphology and color, it is crucial to determine which of these secondary dependencies is more fundamental. Does the halo mass dependence on $n$ arise due to color, or is it the other way around?

Therefore, we select galaxies with rest-frame $u-r$ colors in a narrow range of 2.3 to 2.7. We divide them into three subsamples according to the S\'{e}rsic index, $0.8<n<3$, $3<n<5$ and $5<n<8$. To eliminate the correlation between color and $n$, we resample their color distributions to make them similar by randomly removing some galaxies. The left panel of Figure \ref{fig:fixed} shows their ESD measurements. The red empty squares with errorbars in Figure~\ref{fig:nmc} show the best-fit $M_h$ and $c$ versus $n$, and the ``Fixed color'' row in Table \ref{tab:baseres} presents the best-fit parameters and associated errors. We find that, after fixing the color, the dependencies of both $M_h$ and $c$ on $n$ are still consistent with, and even stronger than,  the results for the blue dots. The best-fit relations are
\begin{equation}\label{eq:kxm_color}
{\rm log}M_h=0.16_{-0.05}^{+0.05} n + 12.81_{-0.25}^{+0.26}\quad[{\rm Fixed \; color}],
\end{equation}
\begin{equation}\label{eq:kxc_color}
c=1.2_{-1.9}^{+1.7} n + 4.5_{-7.1}^{+8.8}\quad[{\rm Fixed \; color}],
\end{equation}
This implies that the correlation between $n$ and halo properties is not predominantly driven by differences in galaxy color. 

Conversely, to explore whether the color dependence is driven by morphology, we also select galaxies with $1 < n < 6$ and divide them into two color subsamples: blue galaxies with $2 < u-r < 2.25$ and red galaxies with $2.25 < u-r < 2.6$. For fair comparison, these two sub-samples are resampled by randomly selecting galaxies to ensure a similar distribution of $n$, with average $n$ of $2.98 \pm 1.07$ and $3.10 \pm 1.09$ respectively, where the errors are the standard deviations of $n$. The middle panel of Figure \ref{fig:fixed} shows their ESD measurements. The orange and blue rings in Figure \ref{fig:nmc} represent the results for the red and blue galaxy samples, with their sizes indicating the 1$\sigma$ error range. The ``Fixed $n$'' section in Table \ref{tab:baseres} shows the best-fit results. We find that even when \(n\) is fixed, halo mass still exhibits a color bimodality, with red galaxies residing in more massive halos than blue galaxies, while their halo concentrations remain consistent. This suggests that a better secondary proxy for halo mass should involve a combination of both color and morphology. Due to data limitations, we are unable to determine the optimal combination, but this presents an intriguing topic for future research.

\subsection{Halo mass dependence on group richness or satellite distribution}
\begin{figure*}[ht!]
\centering
\includegraphics[width=1\textwidth]{ 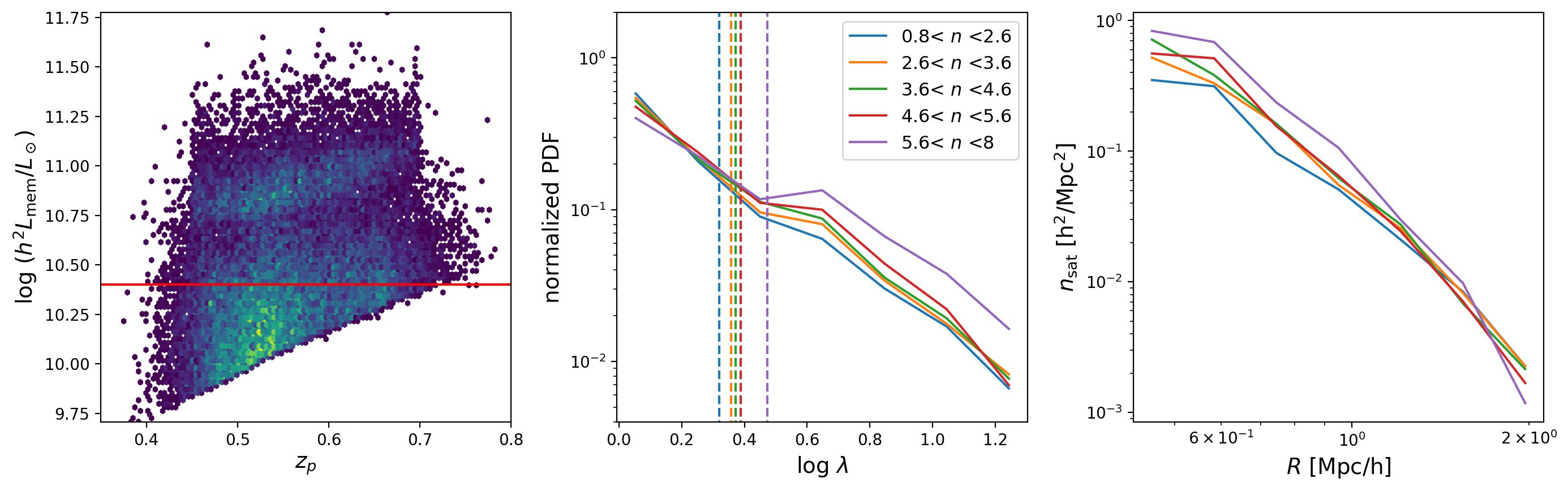}
\caption{The left panel displays the luminosity vs. photo-$z$ diagram of all member galaxies in clusters, with the red solid line indicating the minimum luminosity threshold used to select the member galaxies for the middle and right panels. The middle panel shows the normalized richness distribution of galaxy clusters with different galaxy S\'{e}rsic indices, where dashed lines represent the average richness of each subsample. The right panel displays the projected number density profiles of companion galaxies as a function of projected radius.} \label{fig:yang}
\end{figure*}

Galaxy properties and evolution are typically shaped by their surrounding environment, including local matter density and the evolutionary history of their host halos. We further investigate the properties of galaxy groups hosting the central galaxies using the group catalog from \cite{yang2021extended}, which provides details on group richness $\lambda$ and member galaxies. However, galaxy/group samples from observations are generally flux-limited. As shown in the left panel of Figure \ref{fig:yang}, which plots the luminosity versus photo-$z$ diagram of all member galaxies in groups, the member galaxies and their richness are influenced by different absolute magnitude limits at various redshifts. To mitigate the effects of redshift dependence, we define a volume-limited sample by setting a minimum luminosity threshold for the member galaxies. Consequently, the richness here is defined as the number of member galaxies in a group with ${\rm log}(h^2 L_{\rm men}/L_\odot) > 10.4$.

The middle panel of Figure \ref{fig:yang} shows the richness distributions of groups with different $n$ of central galaxies in our sample. The vertical dashed lines indicate the mean richness for each sample. We find that central galaxies with larger $n$ are more likely to be located in groups with higher richness. Also, we calculate the projected number density profiles of satellite galaxies at various projected radii, displayed in the right panel of Figure \ref{fig:yang}. It shows that the surrounding density of companion satellite galaxies increases with the \(n\) of central galaxies, consistent with the results from \cite{xu2022photometric}. In fact, many previous studies have pointed out the tight correlation between the host halo mass and cluster richness or satellite abundance \citep[e.g.][]{2009ApJ...699..768R,2010MNRAS.404.1922A,2012MNRAS.424.2574W,2017MNRAS.466.3103S,2018ApJ...854..120M,2019PASJ...71..107M,2020MNRAS.491.1643P,2020MNRAS.498.2030C,2021ApJ...919...25W,2021MNRAS.505.5370T,2024ApJ...966..227C}, and thus our results here are related to the fact that galaxies with larger S\'{e}rsic indices are hosted by more massive dark matter halos.

\subsection{Interpretations on the correlation between halo concentration and galaxy S\'{e}rsic index}

In Figure \ref{fig:nmc}, we find a slight positive correlation between the Sérsic index and halo concentration. Also, the right panel of Figure \ref{fig:yang} shows that for galaxy clusters with smaller $n$, the distribution of satellite galaxies is slightly more extended. While more precise measurements are needed to fully confirm this relation, it would be intriguing to explore its theoretical implications.

Numerical simulations predict the trend of lower concentrations in more massive halos \citep[e.g.][]{maccio2007concentration,duffy2008dark,mandelbaum2008halo,correa2015accretion,ludlow2014mass,Diemer2015,ludlow2016mass}, which can be explained by that these halos form later, allowing more material and subhalos to continuously accrete from the surrounding regions, resulting in a more diffuse distribution \citep[e.g.][]{FH21,Gao23, Diemer22, 2024arXiv240714827H}. However, in our results, we find that both the host halo mass and halo concentration are positively correlated with the Sérsic index of galaxies (showing positive $n-M_h$ and $n-c$ relations). If the $n-c$ relation were solely driven by the negative halo mass-concentration relation for dark matter halos, we would expect a negative $n-c$ relation instead. Therefore, the positive correlation we observe between halo concentration and the Sérsic index is likely driven by other factors beyond the typical mass-concentration relation influenced by the environment.

\cite{navarro1997universal} propose that older halos, which formed in higher cosmic densities, tend to have larger characteristic densities and concentrations. Using high-resolution N-body simulations, \cite{zhao2009accurate} find that the halo concentration is tightly correlated with the universe age when its progenitor first reaches 4\% of their current mass, i.e. $t/t_{0.04}$ in their work. Therefore, we speculate that two main factors may influence the $n-c$ relation in our results. One is introduced by the halo mass-concentration relation, that the positive $n-M_h$ relation would introduce negative correlations between $n$ and $c$. 
The other is that more concentrated galaxies with larger S\'{e}rsic indices form earlier at fixed stellar mass, with their host halos form earlier as well, and dark matter halos with earlier formation times have larger concentrations. Among these two factors, the latter may play a dominant role for our galaxy samples.

To further test our hypothesis, we control halo mass in the sample by using the halo mass from the group catalog \(M_{G}\), which is assigned using the abundance matching technique \citep{yang2021extended}. Considering the positive correlation between the S\'ersic index $n$ and halo mass $M_h$, we could adjust the selection range of $M_G$ to ensure the halo mass is the same for different galaxy samples. Specifically, we divide the sample into two groups based on S\'ersic index: for galaxies with $0.8 < n < 4$, we select samples satisfying $10^{13.3} < M_G < 10^{13.8}\rm M_\odot/h$; for galaxies with $4 < n < 8$, we use samples with $10^{13.3} < M_G < 10^{13.7} M_\odot/h$. The right panel of Figure \ref{fig:fixed} shows the ESD measurements, Table \ref{tab:baseres} presents sample information and best-fit parameters, which are also shown by the green squares in Figure \ref{fig:nmc}. We find that the halo masses from the group catalog of the two subsamples are very consistent, but halo concentrations for galaxies with larger S\'{e}rsic indices are higher, still consistent with the $n-c$ relation shown and discussed earlier. Furthermore, compared to the $n-c$ relation without controlling halo mass, the relation becomes steeper. This implies that by reducing the negative $M_h-c$ relationship caused by accretion and merging, the positive correlation between $n$ and $c$ becomes slightly strengthened. This also supports our speculation that the host halos of galaxies with higher S\'{e}rsic indices are formed earlier, leading to higher halo concentrations, and this effect is stronger than the influence of later-time halo mergers to decrease the concentration at least for the massive CMASS galaxies at $z\sim0.57$.

From the perspective of assembly bias, we find that the halo bias of the galaxy sample with higher S\'ersic index is lower, which is also reflected in the lower large-scale ESD shown in Figure \ref{fig:fixed}. \citet{Wechsler2006ApJ} proposed that for massive halos with fixed halo mass, those form earlier (with higher concentration) have weaker clustering, resulting in a smaller halo bias. From this standpoint, our findings further support the inference that halos hosting galaxies with higher S\'ersic indices formed earlier. This also suggests that the S\'ersic index could serve as a proxy for halo concentration or formation time, which could be used in the future to probe the assembly bias of galaxy clusters.

\subsection{Comparison with other results}
\begin{figure}[ht!]
\centering
\includegraphics[width=0.45\textwidth]{ 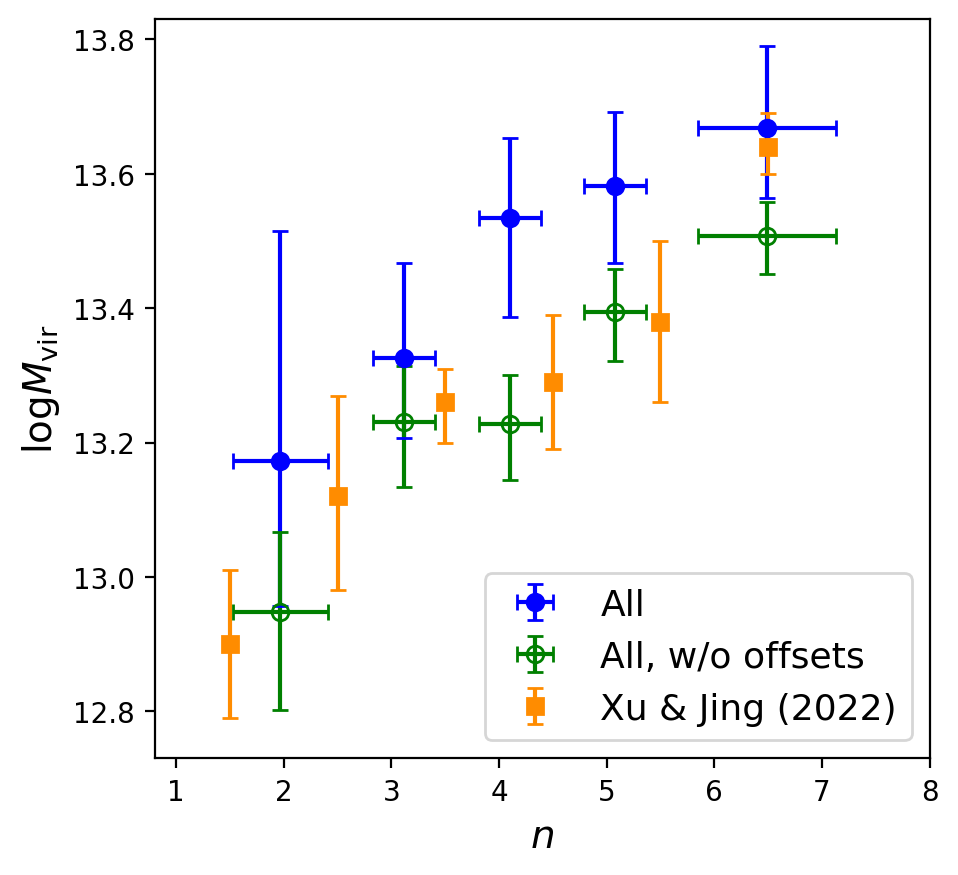}
\caption{The relationships between $n$ and halo mass measured in different ways. The definition of halo mass $M_{\rm vir}$ here is from \cite{Bryan_1998}. Blue dots represent our results for all samples. Green hollow points indicate the results obtained without considering off-centering effects in the halo model. Orange squares display the results from \cite{xu2022photometric}.
\label{fig:nMxu}}
\end{figure}

Figure \ref{fig:nMxu} compares our main results (blue dots) with those obtained using the PAC method in the work of \cite{xu2022photometric} (orange squares).  To ensure a consistent basis for comparison, we convert the halo masses in our results to their definition, where $\Delta_{\rm vir}= 18 \pi^2 + 82 x -39 x^2$ and $x=\Omega_m(z)-1$. We observe that our results systematically yield higher halo masses. This discrepancy may stem from the assumption in \cite{xu2022photometric} that massive central galaxies are perfectly situated at the halo centers, neglecting the off-centering effects in observations. By setting $f_c=1$ in our model, disregarding the central offsets, we show the fitting results as green hollow points in the figure. In this test, our results align well with theirs.

\section{Comparison with Simulations}\label{sec:siml}

So far we have probed how do the host halo properties correlate with galaxy and galaxy cluster properties. However, the physical drivers behind these relationships are still under debates. \cite{cui2021origin} quantitatively reproduce the color bimodality of the SHMR in the SIMBA simulations \citep{dave2019simba}, showing that red galaxies reside in more massive halos at a given stellar mass. They find that this is mainly due to active galactic nuclei (AGN) feedback in the form of jets and X-rays, which significantly suppresses star formation earlier in red galaxies while their halos continue to grow. This also could be the reason of the \( n-M_h \) relationship. Therefore, we further investigate similar relations in this section, using two hydrodynamic simulations, TNG and SIMBA.

\begin{figure*}[ht!]
\centering
\includegraphics[width=1\textwidth]{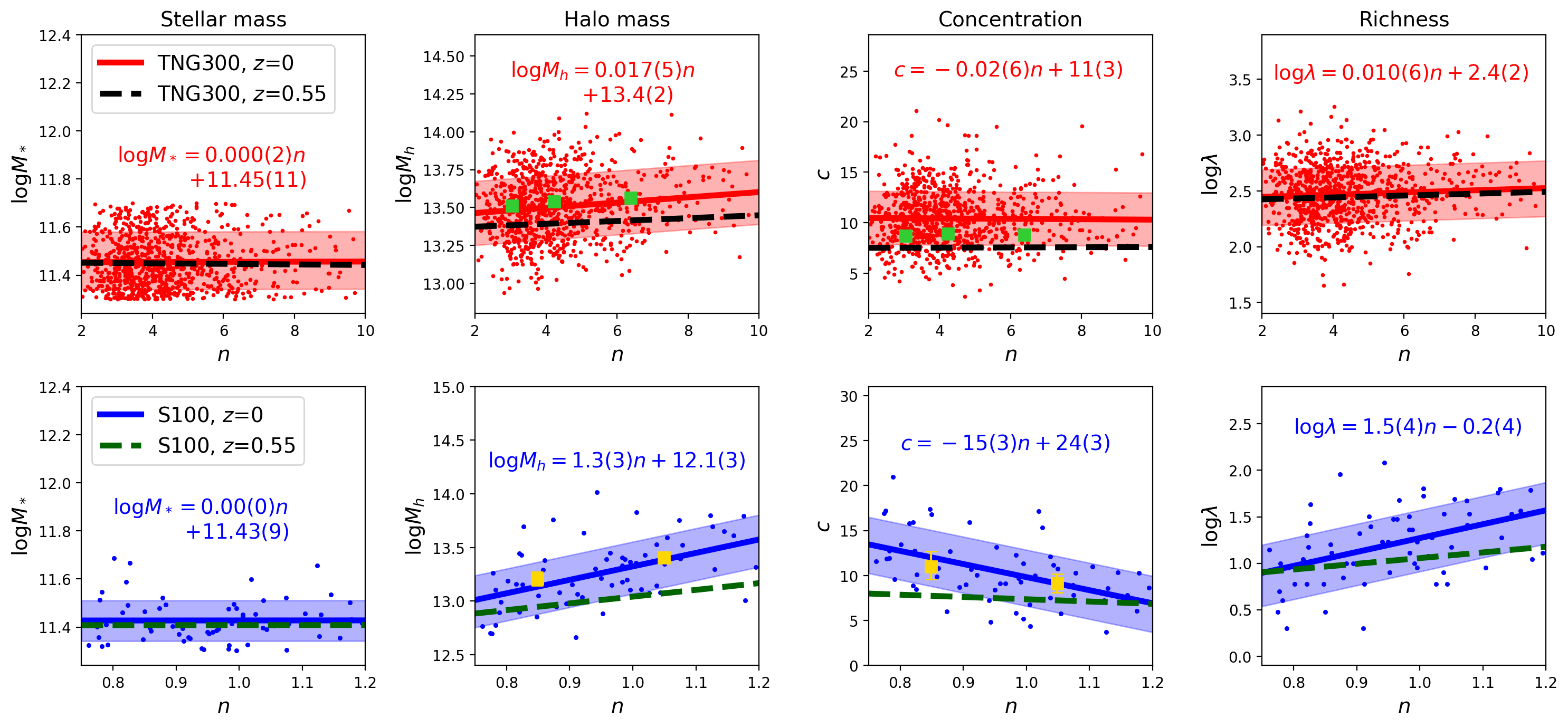}
\caption{The correlations between the ${\rm S \Acute{e}rsic}$ index ($n$) of central galaxies and their host halo mass, halo concentration, and richness in two simulations, S100 and TNG300. Central galaxies are selected to have stellar masses in the range of $11.3<{\rm log}M_*/{\rm M_\odot} < 11.7$ for snapshots at redshift of 0 and 0.55. Solid lines represent linear best-fit results for samples at $z=0$ (dots), with shaded areas indicating the 16th to 84th percentiles. While dash lines represent the fitting results for samples at $z=0.55$. The formula provided in each panel gives the best-fit result to galaxy systems at $z=0$, in which the numbers in the parentheses represent the 1$\sigma$ errors of the last digit or last two digits. The squares in the ${\rm log}M_h - n$ and $c-n$ panels represent the best-fit results from the stacked ESD measurements in Figure \ref{fig:ggl_siml}, where the errorbars are too small to be visible.}\label{fig:mn_siml}
\end{figure*}

\subsection{The relationship between S\'{e}rsic index and halo properties} 
Before proceeding, we need to clarify some problems when measuring the galaxy morphology in SIMBA: the galaxies in these simulations generally exhibit very low galaxy concentrations, with $n$ not exceeding 2. According to \cite{dave2019simba}, the size of low-mass quenched galaxies in SIMBA is a notable failure of the current model, as their half-light radii are larger than those of star-forming galaxies, which contradicts observational trends. A larger radius usually implies a lower concentration, leading to elliptical galaxies that should have higher $n$ but obtaining very low $n$ values in the simulation. Nevertheless, we endeavor to incorporate the results from SIMBA in our analysis within the limited range of \( n \) it produces. 

\begin{figure*}[ht!]
\centering
\includegraphics[width=0.6\textwidth]{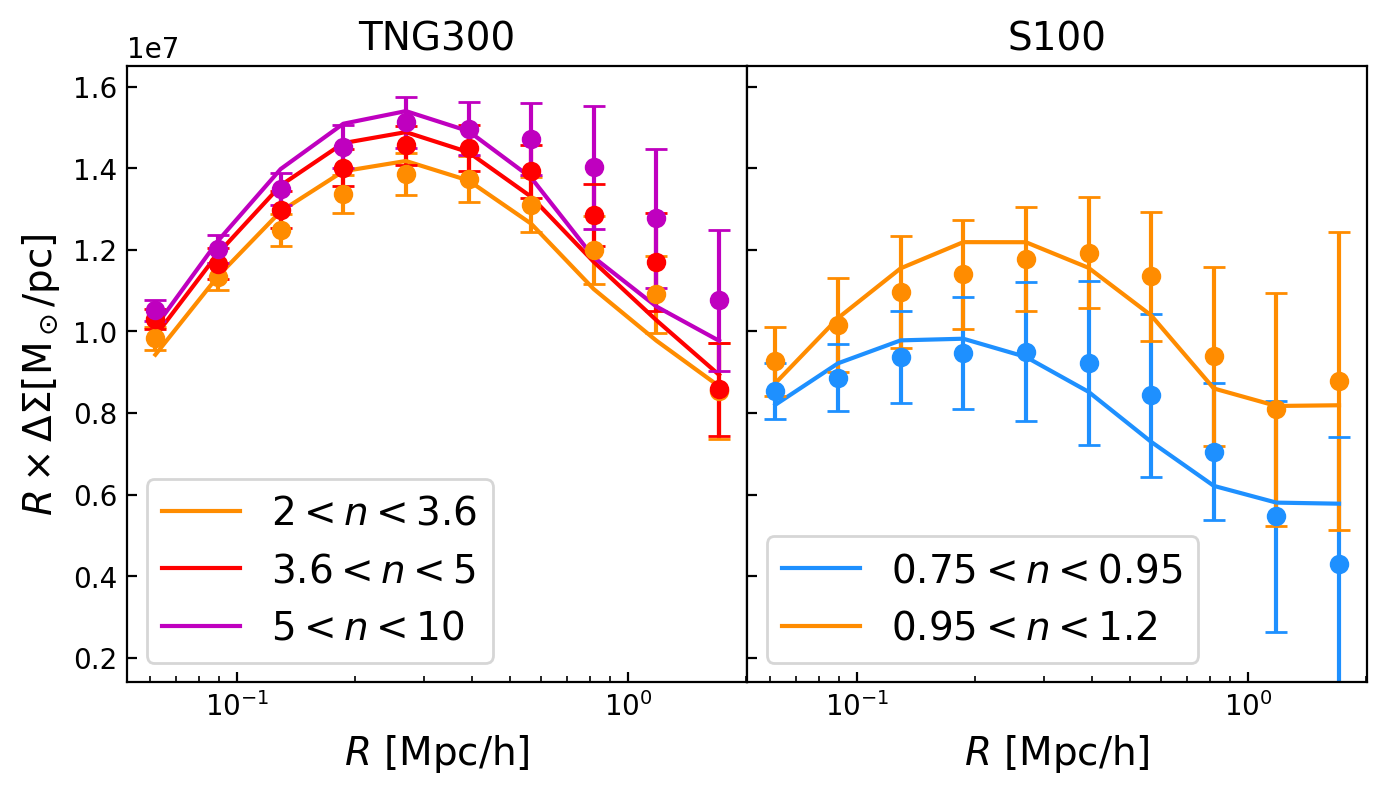}
\caption{Dots with errorbars are stacked ESDs of halos in Figure \ref{fig:mn_siml} within different range of S\'ersic index $n$. Solid lines are the best fits with 3 free parameters: $M_h$, $c$ and $b_h$. The left panel shows the results from TNG300, corresponding to the upper panels of Figure \ref{fig:mn_siml}, while the right panel is for S100, corresponding to the lower panels of Figure \ref{fig:mn_siml}.}
\label{fig:ggl_siml}
\end{figure*}
Firstly, we use snapshots from TNG300 and SIMBA at two redshifts for the main analysis, $z=0$ and $z=0.55$ respectively\footnote{The latter is actually $z=0.556839$ for SIMBA, but denoted as 0.55.}. Figure \ref{fig:mn_siml} displays the relationships between the ${\rm S \Acute{e}rsic}$ index, \( n \), and the halo mass, \( M_h \), halo concentration, \( c \), and richness, \( \lambda \), for central galaxies in simulations. The properties of galaxies and halos are represented by scatter plots, with the linear fits represented by lines, and the 16-84 percentiles displayed as shaded areas. Note that, since there is a slight correlation between $n$ and $M_*$, we randomly remove galaxies at different $n$ to decorrelate these two parameters. As a result, the slope of the relationship in the leftmost panel of Figure \ref{fig:mn_siml} is zero, thereby avoiding the impact of stellar mass on other correlations between $n$ and halo properties.

We also bin the galaxies into 3 groups in TNG300 and 2 groups in S100, and stacked their ESDs as in the observations. The results are presented in Figure \ref{fig:ggl_siml} with their best-fit curves. Similar to fitting individual halos, we assume that the central galaxies in the simulations are well-centered and treat \(M_h\), \(c\), and \(b_h\) as free parameters for fitting. The fitting results for \(M_h\) and \(c\) are shown as green and yellow squares in Figure \ref{fig:mn_siml}. Note that the error in the stacked fitting results represents the uncertainty of the average properties of halos, which is so small that it is almost invisible in the figure. The stacked fitting results are close to the best-fit solid lines of scatter points obtained from fitting individual halos, demonstrating the robustness of the stacking lensing measurements.

Regarding the relationship between halo mass and $n$, we find that both TNG300 and SIMBA exhibit correlations,  also shown in the rising ESD in Figure \ref{fig:ggl_siml}. However, compared to observational results, halo masses from TNG300 are higher and their correlation with $n$ is weaker (Eq.\ref{eq:kxm_all}). While the S100 simulation shows a stronger correlation, considering its incorrect galaxy morphology, we only qualitatively analyze the trends. Both simulations demonstrate a positive correlation between $n$ and halo abundance $\lambda$, which is qualitatively consistent with our results (Figure \ref{fig:yang}). Moreover, we note that the correlations between $n$ and other galaxy and halo properties at higher redshift are weaker than those at lower redshift.

\begin{figure*}[ht!]
\centering
\includegraphics[width=0.9\textwidth]{ 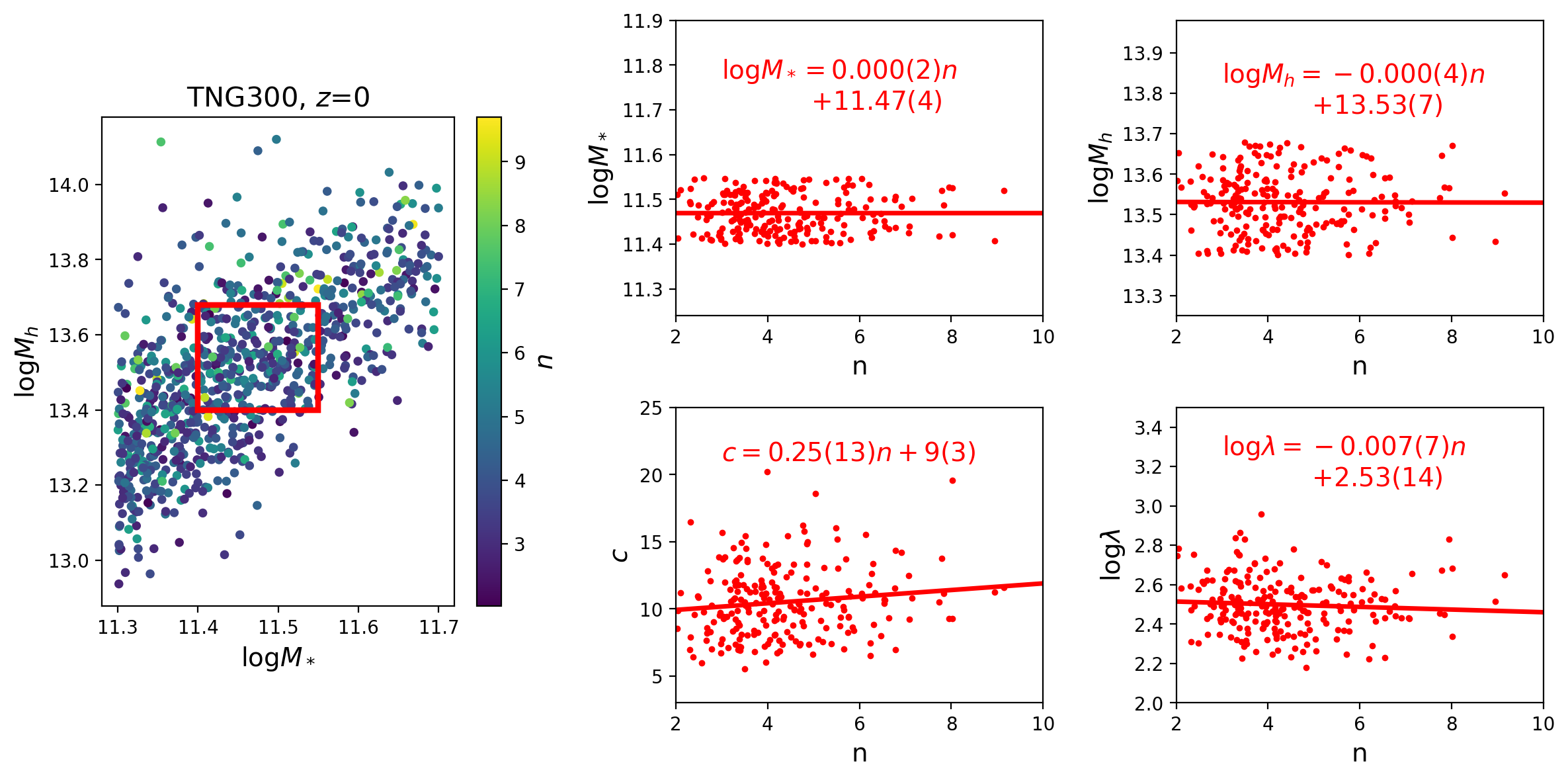}
\caption{The left panel shows halo mass ($M_h$) as a function of stellar mass ($M_\ast$) for samples in the TNG300 simulation at $z=0$ (red dots in Figure \ref{fig:mn_siml}), colour-coded by their S\'{e}rsic indices, $n$. We select the subsample within the red box to simultaneously control both $M_*$ and $M_h$. The right panel displays the dependencies of the S\'{e}rsic index, $n$, on stellar mass, halo mass, halo concentration, and richness for this subsample.\label{fig:fixm_siml}}
\end{figure*}

However, neither of the two simulations reproduce significant positive correlations between halo concentration and $n$ as have been observed in the real data. A strong negative correlation  even shows in SIMBA. As previously discussed, the $n-c$ relation at fixed stellar mass is affected by: 1) the mass-concentration  ($M_h-c$) relation of dark matter halos; and 2) the intrinsic $n-c$ relation at fixed halo mass. To check whether the TNG simulation reveals a similar trend between $n$ and $c$ as what we have inferred from the data, we control both halo mass and stellar mass in a subsample from the TNG300 at $z=0$. The left panel of Figure \ref{fig:fixm_siml} shows the correlation between halo mass and stellar mass for the red points in Figure \ref{fig:mn_siml}, color coded by $n$. We select galaxies within the red box in the panel and replot the correlations between $n$ and other properties for them in right panels of Figure \ref{fig:fixm_siml}. We find that after controlling both the stellar mass and halo mass to a narrow range, the halo concentration shows a slight positive correlation with $n$. This might imply that, for central galaxies with the same stellar mass and halo mass, more concentrated central galaxies are more likely to reside in more concentrated halos, which is consistent with our previous hypothesis and results. However, the correlation is still weak and may depend on the details of galaxy formation models adopted in the simulations and how realistic are these models.

\subsection{The effects of AGN feedback in SIMBA}
\begin{figure*}[ht!]
\centering
\includegraphics[width=1\textwidth]{ 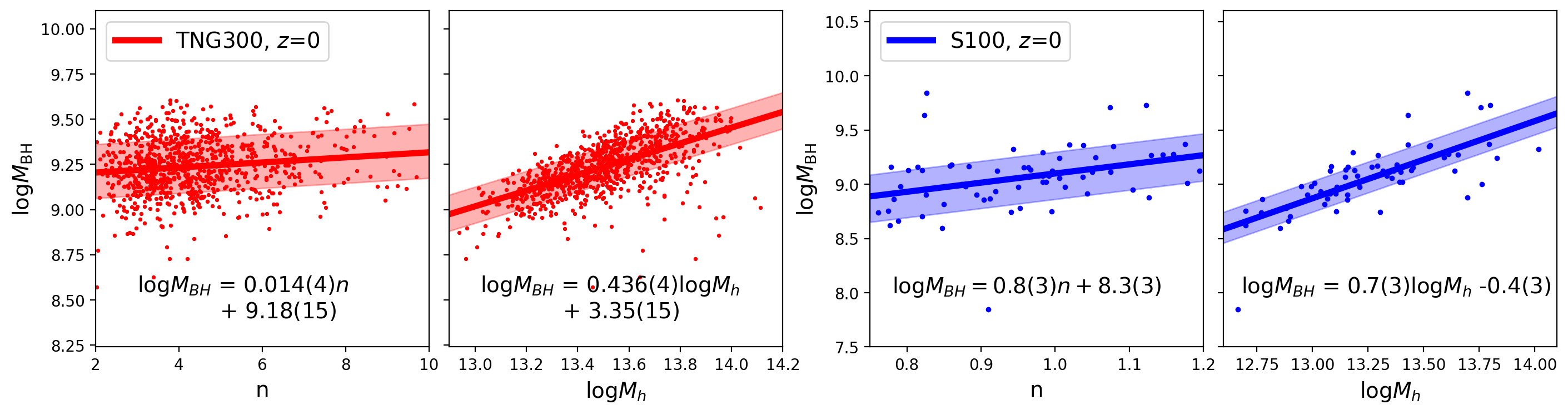}
\caption{The relationship among $n$, black hole mass $M_{\rm BH}$ and halo mass $M_h$ for galaxies from
S100 and TNG300 at $z=0$ (corresponding to the red and blue dots in Figure \ref{fig:mn_siml}). The solid lines represent linear best-fit results and shaded areas represent the
16th to 84th percentiles, and the formulas in panels are equations of the solid lines.
\label{fig:BH_siml}}
\end{figure*}

We further examine the relationship between $n$ and the black hole (BH) mass $M_{\rm BH}$ in the galaxies for the two simulations at $z=0$, as shown in Figure \ref{fig:BH_siml}. We find a weak positive correlation between $n$ and $M_{\rm BH}$ in both TNG300 and S100. It might be because a larger $n$ typically implies that a galaxy has a more dominant bulge component, and many studies reported a positive correlation between bulge and black hole mass \citep{mclure2002black, wandel2002black, mcconnell2013revisiting, davis2019black}. We also display the relationship between $M_{\rm BH}$ and $M_h$, revealing strong positive correlations. We can derive the relationship between ${\rm log}M_h$ and $n$ by combining the best-fit equations in ${\rm log}M_{\rm BH}-n$ and ${\rm log}M_{\rm BH}-{\rm log}M_h$ panels, and we find that the results are consistent with those obtained from the direct fitting shown in Figure \ref{fig:mn_siml}. This implies that the correlation between the ${\rm S \Acute{e}rsic}$ index and halo mass might be related to the different black hole activities.

\begin{figure*}[ht!]
\centering
\includegraphics[width=1\textwidth]{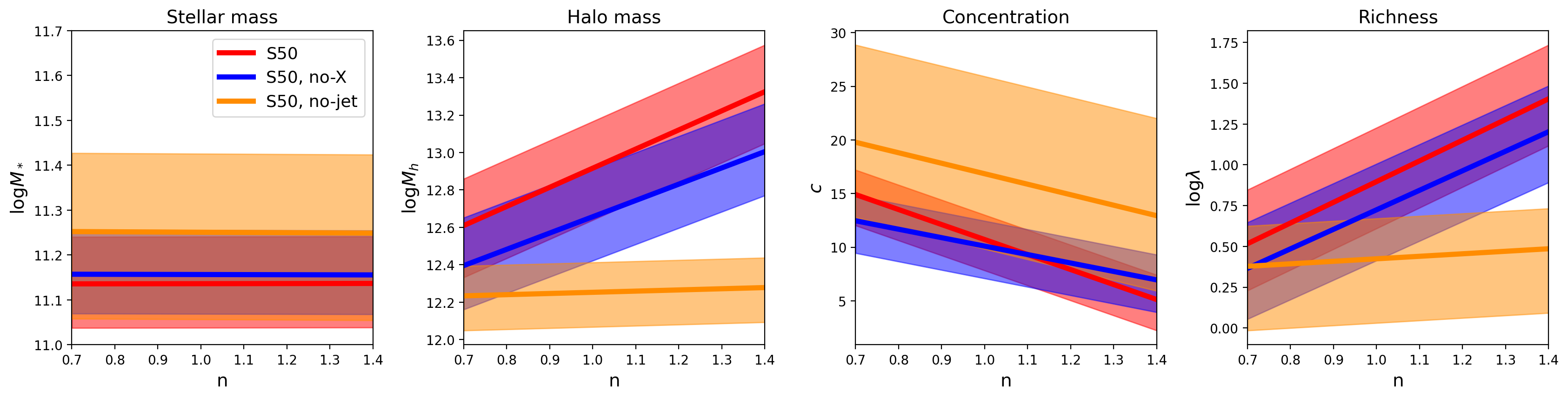}
\caption{Similar to Figure \ref{fig:mn_siml}, but the results are from three different simulations. ``S50'' represents the SIMBA simulation with a box length of 50 Mpc/$h$. ``no-X'' indicates a simulation without X-ray feedback, and ``no-jet'' represents a simulation without X-ray and jet mode AGN feedback.
\label{fig:AGN_siml}}
\end{figure*}

According to \cite{thomas2019black}, the Eddington ratio, $f_{\rm Edd}$, of black holes is inversely correlated with their mass. When $f_{\rm Edd} < 0.2$, it would trigger jet-mode AGN feedback and subsequent X-ray AGN feedback. Therefore, to validate the $n-M_h$ relationship in SIMBA influenced by black hole mass, we use simulations from the SIMBA series with different modes of AGN feedback, including simulations that incorporate all physical processes (``S50''), that exclude X-ray feedback (``S50, no-X''), and that exclude both X-ray feedback and jet-mode AGN feedback (``S50, no-jet''). Figure \ref{fig:AGN_siml} shows the correlations between $n$ of central galaxies and other halo properties for three simulations, where the shaded regions represent the 16-84 percentiles. We observe distinctly different trends of $n$ with other properties in the three simulations. As shown in the leftmost panel, the average stellar mass of galaxies without X-ray and jet-mode AGN feedback (orange) is higher than those with jet-mode AGN feedback (red and blue). This indicates that jet-mode feedback primarily reduces the star formation rate in galaxies, as stated in \cite{dave2019simba}. Furthermore, as AGN feedback modes are progressively turned off, 
the slope between halo mass and $n$ gradually decreases.
Especially in the ``S50, no-jet'' simulation, the correlation between $n$ and $M_h$ almost disappears, and the correlations with $c$ and $\lambda$ are also significantly weakened. These results indicate that the relationship between $n$ and halo properties might be influenced by the strength of AGN feedback, especially the jet mode, which can rapidly and effectively expel gas from galaxies, leading to the cessation of star formation.

\section{Summary and conclusion}\label{sec:conclusion}

We delve into the relationship between the morphology of central galaxies indicated by S\'ersic index, $n$, and the properties of their host dark matter halos and galaxy groups/clusters, including halo mass, halo concentration and group richness. Our study in this paper extends the work of \cite{xu2022photometric}, in which the morphology, color, and size dependencies of the SHMR for massive central galaxies have been investigated using the PAC method. We use the central galaxy sample from their work, with a mass range of $11.3 < {\rm log}(M_*/{\rm M_\odot}) < 11.7$ and a redshift range of $0.45-0.7$, selected from BOSS CMASS observations. We employ galaxy-galaxy lensing to measure the ESD around these galaxies, using the Fourier\_Quad shear catalog from the third public data release of HSC. In our modeling approach, halo mass \( M_h \) and concentration \( c \) from the NFW profile were considered as free parameters, while also incorporating off-centering effects and 2-halo terms. We find the following:
\begin{itemize}
\item From lensing measurements, we discover a strong positive correlation between halo mass and the S\'{e}rsic index $n$ of central galaxies with fixed stellar mass, demonstrating that more concentrated central galaxies reside in more massive halos.

\item To investigate whether galaxy S\'{e}rsic index, $n$, or galaxy color is a more fundamental secondary properties in determining halo mass, we separately control $n$ and color for galaxies to be in a narrow range. The dependence of halo mass on both $n$ and galaxy color still exists, which might indicate that some combinations of color and $n$ can be a better secondary proxy.

\item We find a slight positive correlation between $n$ and halo concentration $c$ with postive $n-M_h$ relations. In contrast, the simulation results show that more massive halos are less concentrated due to their more frequent mergers at late universe. After controlling for both stellar mass and halo mass, we observe a slight strengthening of the $n-c$ dependence. This suggest that the $n$ of galaxies may serve as an indicator for the secondary $M_h-c$ relation, which contributes to the scatter of general $M_h-c$ relation in simulations. We therefore infer that, at fixed stellar mass, more concentrated galaxies and their host halos form earlier and higher concentrations.

\item Using the group catalog from \cite{yang2021extended}, we find that more concentrated central galaxies are located in richer galaxy clusters and have a higher average satellite galaxy number density around them.

\item TNG300 and SIMBA simulations qualitatively reproduce the positive correlation between $n$ and halo mass as well as richness, but the correlation in TNG300 is very weak. Additionally, neither simulation shows positive $n-c$ relationships. By further controlling halo mass in the TNG300 sample, we find a weak positive $n-c$ correlation.

\item In TNG300 and SIMBA, there are slight positive correlations between $n$ and black hole mass, and strong positive correlations between black hole mass and halo mass. As we progressively turn off X-ray feedback and jet-mode AGN feedback in SIMBA simulations, the correlation between $n$ and halo mass gradually disappeared. This suggests that AGN feedback might play a role in the halo mass dependence on secondary galaxy properties mentioned above.
\end{itemize}

Our detection of the halo mass dependence on secondary galaxy properties bring better insights into the physics of galaxy assembly bias, which is an important problem in the current galaxy-halo connection field. With next-generation surveys and better measurements, the connection between galaxies and halos should be better constrained.

\section*{Acknowledgements}
This work is supported by the National Key Basic Research and Development Program of China (2023YFA1607800, 2023YFA1607802), the NSFC grants (11621303, 11890691, 12073017), and the science research grants from China Manned Space Project (No. CMS-CSST-2021-A01). Z.L. is supported by the funding from China Scholarship Council. K.X. is supported by the funding from the Center for Particle Cosmology at U Penn. W.W. is supported by NSFC (12273021) and the National Key R\&D Program of
China (2023YFA1605600, 2023YFA1605601). The computations in this paper were run on the $\pi$ 2.0 cluster supported by the Center of High Performance Computing at Shanghai Jiaotong University, and the Gravity supercomputer of
the Astronomy Department, Shanghai Jiaotong University.

The Hyper Suprime-Cam (HSC) collaboration includes the astronomical communities of Japan and Taiwan, and Princeton University. The HSC instrumentation and software were developed by the National Astronomical Observatory of Japan (NAOJ), the Kavli Institute for the Physics and Mathematics of the Universe (Kavli IPMU), the University of Tokyo, the High Energy Accelerator Research Organization (KEK), the Academia Sinica Institute for Astronomy and Astrophysics in Taiwan (ASIAA), and Princeton University. Funding was contributed by the FIRST program from Japanese Cabinet Office, the Ministry of Education, Culture, Sports, Science and Technology (MEXT), the Japan Society for the Promotion of Science (JSPS), Japan Science and Technology Agency (JST), the Toray Science Foundation, NAOJ, Kavli IPMU, KEK, ASIAA, and Princeton University.

This publication has made use of data products from the Sloan Digital Sky Survey (SDSS). Funding for SDSS and SDSS-II has been provided by the Alfred P. Sloan Foundation, the Participating Institutions, the National Science Foundation, the U.S. Department of Energy, the National Aeronautics and Space Administration, the Japanese Monbukagakusho, the Max Planck Society, and the Higher Education Funding Council for England.

\appendix
\section{Field Distortion Test}\label{sec:FD}
\begin{figure}[ht!]
\centering
\includegraphics[width=0.4\textwidth]{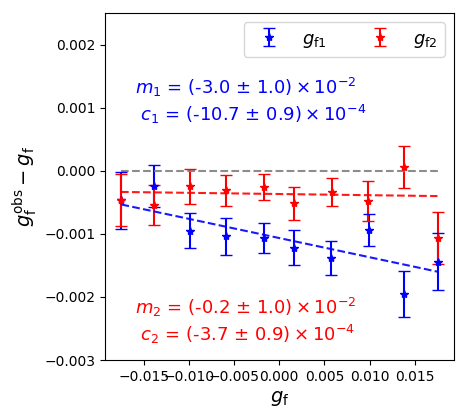}
\caption{Field distortion test. $g_f$ represents the field distortion values from the shear catalog, while the vertical axis represents the difference between measured values $g_f^{\rm obs}$ and true values. Blue and red correspond to the two components of the shear. $m_i$ and $c_i$ represent the multiplicative and additive biases obtained from linear fitting, respectively. \label{fig:FD}}
\end{figure}

Here, we measure the multiplicative biases introduced in shear measurements and correct it in our analysis. Such biases arise from systematic errors or the intrinsic noisiness of the shear estimators, and they are typically quantified linearly,
\begin{equation}\label{eq:gf}
    g^{\rm obs}=(1+m_{\rm bias})g^{\rm true}+c_{\rm bias},
\end{equation}
where $m_{\rm bias}$ is multiplicative bias and $c_{\rm bias}$ is addictive bias, $g^{\rm true}$ refers to the real shear and $g^{\rm obs}$ represents the measured shear. 

In an ideal scenario, astronomical observation equipment should provide distortion-free images. However, the images are always distorted due to imperfections in the optical system or imaging equipment, known as field distortion (FD hereafter). This distortion affects the shapes and the positions of galaxies and stars, introducing an equivalent shear, with two components denoted as \( g_{f1} \) and \( g_{f2} \). It is typically related to the object’s position on the focal plane rather than its position in the sky. \cite{Zhang2019ApJ} proposed using FD to detect \( c_{\rm bias} \) and \( m_{\rm bias} \) in shear measurements, which is a calibration method based on the properties of the data itself. The two components of FD signals of each galaxy, \( g_{f1} \) and \( g_{f2} \), are provided in the FQ shear catalog. By stacking the galaxy shear estimators at fixed FD values, the cosmic shear signal will vanish since these galaxies are from different positions on the sky, allowing us to recover the FD value. By comparing the measured FD signals with the true FD signals from the catalog and using Eq. \ref{eq:gf}, the shear bias in the measurements can be determined. 
 
We perform a FD test on all background galaxies in the ESD measurements, and we weight each background galaxy by $\Sigma_c$ (Eq. \ref{sigmac}) to better match the ESD scenario. Since we use background galaxies with $|g_{fi}| < 0.02 (i=1,2)$, we divide $g_{fi}$ evenly into 10 bins from -0.02 to 0.02, and stack the galaxy shear estimators within each bin to calculate the observed shear value $g_f^{\rm obs}$. Figure \ref{fig:FD} displays the measured signals for the two shear components, where \( g_f \) is the true FD value, and the vertical axis represents the difference between the observed distortion value \( g_f^{\rm obs} \) and the true value. If there is no shear measurement bias, the results in the plot should all align with 0, i.e., $g_f^{\rm obs}=g_f$. The figure also shows the shear bias results obtained from fitting, revealing that the $g_1$ component is underestimated by about 3\%. Consequently, we correct the shear estimators $G_i$ ($i=1,2$) in the catalog as $G_i/(1+m_i)$ before performing galaxy-galaxy measurements in Eq. \ref{Ghat}. Additionally, $c_{\rm bias}$ is largely mitigated by rotating background galaxies toward the lenses during galaxy-galaxy lensing calculations, with further reduction through the subtraction of random point signals.

\section{The ESD model of galaxy}\label{app:sersic}
\begin{figure}[ht!]
\centering
\includegraphics[width=0.4\textwidth]{ 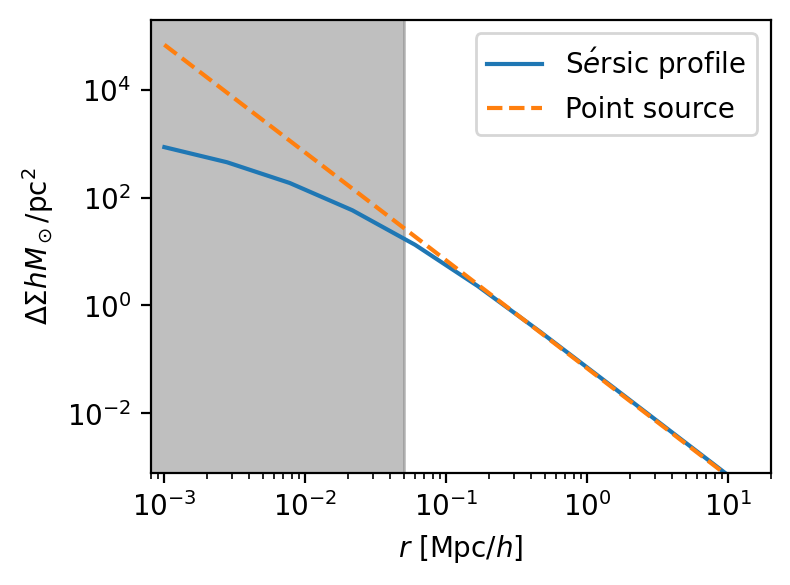}
\caption{The ESD of galaxy using S\'ersic profile and point source model. The white region represents the radius range in our measurements.} \label{fig:point}
\end{figure}

In this study, we assume that galaxies can be treated as point sources for measuring the ESD. The purpose of this section is to show that, within the radial range measured in this work, using point sources does not significantly affect the ESD results. Typically, assuming a constant mass-to-light ratio within galaxies, their light distribution can be considered the same as their mass distribution. A more accurate ESD model would involve using the S\'ersic profile from Eq. \ref{eq:sersic} in Eq. \ref{eq:ESD} to calculate the ESD. However, when measuring the ESD of halos, galaxies are often treated as very small compared to the halos and are approximated as point sources, meaning their density profile is a Dirac delta function \(\delta(r)\). Since the minimum radius for our ESD measurement is set at 0.05 Mpc/$h$, which is comparable to the size of galaxies. Therefore, we need to verify whether the point source model is valid at this scale. In the TNG300 simulation described in Section \ref{sec:siml}, the average half-light radius \(R_e\) of the galaxies shown in Figure \ref{fig:mn_siml} is about 20 kpc/$h$. Using this, we create a simple S\'ersic profile and set the galaxy mass to \(10^{11.5} M_\odot\). Figure \ref{fig:point} shows the ESD results for two models. It is clear that within the white region (the radial range in our measurement), the ESD results from both models are almost identical, but point source model would overestimate at smaller scales (gray region). Thus, within the current radial range, using the point source model for simplified calculations does not affect the reliability of the conclusions in this paper.

\bibliography{sample631}{}
\bibliographystyle{aasjournal}



\end{document}